\documentclass[twocolumn,showpacs,floatfix]{revtex4-1}
\usepackage[utf8]{inputenc}
\usepackage{bm,amsmath,amssymb,mathtools,color,graphicx}
\usepackage{textcomp,gensymb,units,grffile}
\usepackage[a4paper]{geometry}

\usepackage[pdftex]{hyperref}

\newcommand{\xb}{{\bar x}}
\newcommand{\be}{\begin{equation}}
\newcommand{\ee}{\end{equation}}
\newcommand{\bea}{\begin{eqnarray}}
\newcommand{\eea}{\end{eqnarray}}

\begin{document}

\title{Selection dynamics in transient compartmentalization}
\author{Alex Blokhuis}
\author{David Lacoste}
\affiliation{Gulliver Laboratory, UMR CNRS 7083, PSL Research University, 
ESPCI, 10 rue Vauquelin, F-75231 Paris Cedex 05, France}
\author{Philippe
 Nghe}
\affiliation{Laboratory of Biochemistry, PSL Research University, ESPCI, 10 rue Vauquelin, F-75231 Paris Cedex 05, France}
\author{Luca Peliti}
\email{luca@peliti.org}
\affiliation{SMRI, 00058 Santa Marinella (RM), Italy}
\date{\today} 

\begin{abstract}
Transient compartments have been recently shown to be able to 
maintain functional replicators in the context of prebiotic studies.
Motivated by this experiment, we show that a broad class of selection dynamics is
able to achieve this goal. We identify two key parameters, 
the relative amplification of non-active replicators (parasites) and 
the size of compartments. Since the basic ingredients of our model are the competition between a host and its parasite, 
and the diversity generated by small size compartments,
our results are relevant  
to various phage-bacteria or virus-host ecology problems.

\end{abstract}


\maketitle

A central issue in origin of life studies is to explain how replicating functional molecules could have appeared and evolved 
towards higher complexity \cite{Higgs2015}.
In 1965, Spiegelman showed experimentally that RNA could be replicated by an enzyme called $Q\beta$ RNA replicase, 
in the presence of free nucleotides and salt. Interestingly, he noticed that as the process is repeated, shorter and shorter
RNA polymers appear, which he called parasites. Typically, these parasites are non-functional molecules which replicate faster than the 
RNA polymers introduced at the beginning of the experiment and which for this reason tend to dominate. 
Eventually, a polymer of only 218 bases remained out of the original chain of 4500 bases, which became known as Spiegelman's monster.
In 1971, Eigen conceptualized this observation by showing that for a given accuracy of replication and relative fitness of parasites, 
there is a maximal genome length that can be maintained without errors \cite{Eigen1971}. This result led to the following 
paradox: to be a functional replicator, a molecule must be long enough. However, if it is long, it can not be maintained since  
it will quickly be overtaken by parasites. 
Many works attempted to address the puzzle as reviewed in Ref.~\cite{Takeuchi2012}. 
In some recent studies, spatial clustering was found to promote the survival of cooperating 
replicators \cite{Levin2017,Tupper2017, Kim2016}.
This kind of observation is compatible with early theoretical views~\cite{Szathmary1987,Smith1995}, that compartmentalization could allow parasites to be controlled. 

Small compartments are ideal for prebiotic scenarios, because they function as micro-reactors where chemical reactions are facilitated. 
Oparin imagined liquid-like compartments called coacervates, which could play a central role in the origin of life \cite{Oparin1952}. 
Although experimental verification of the prebiotic relevance of coacervates or other sorts of protocells remained 
scarce for many years, the idea has resurfaced recently in various systems of biological interest \cite{Zwicker2017,Brangwynne2009}. 
An important aspect of the original Oparin scenario which has not been addressed in these studies is the possibility of a transient nature of the compartmentalization. In the present paper, we revisit group selection with transient compartmentalization. 
We were motivated by the relevance of transient compartmentalization in several 
scenarios~\cite{hydrovent,vesicles,clusters,Damer2015,Baaske2007} for the origin of life and 
by a recent experiment, in which small droplets 
containing RNA in a microfluidic device \cite{Matsumura2016} were used as compartments.
In this experiment, cycles of transient compartmentalization
 prevent the takeover by parasitic mutants.
Cycles consist of the following steps: (i) inoculation, in which droplets are inoculated with 
a mixture of RNA molecules containing active ribozymes and inactive parasites, (ii) maturation, in which RNA is replicated by $Q \beta$ replicase,  
(iii) selection, in which compartments with a preferred value of the catalytic activity are selected, 
(iv) pooling, in which the content of the selected compartments is pooled. This protocol does not correspond to that 
of the Stochastic Corrector model~\cite{Szathmary1987} because of step~(iv), which removes the separation between individual compartments.

The absence of parasite takeover was successfully explained in ref.~\cite{Matsumura2016}
by a theoretical model which described the appearance of parasites within 
a given lineage as a result of mutations during the replication process. 
In this work we wish to account for these observations in a more general sense. 
We show that the value of the mutation rate does not play an essential 
role as long as it is small \cite{SM}, 
and that the entire shape of the selection function 
is not needed to describe the fate of the system.

Let us consider an infinite population of compartments. Each compartment is 
initially seeded with $n$ replicating molecules, where $n$ is a random
variable, Poisson distributed with average equal to $\lambda$. 
In addition, each compartment also contains a large and constant  
numbers of enzymes, $n_{Q \beta}$ and of activated nucleotides $n_u$.
Among the $n$ replicating molecules, 
$m$ are ribozymes, and the remaining $n-m$ are parasites.
Let $x$ be the initial fraction of ribozymes and $1-x$ that of parasites. After this 
inoculation phase, compartments evolve by letting the total number of molecules 
grow by consuming activated nucleotides. 

In practice, the time of incubation of the compartments is fixed and longer than the time after 
which activated nucleotides become exhausted. 
The kinetics is initially exponential because the synthesis of 
RNA is autocatalytic at low concentration of templates. 
Therefore, the average number  
${\bar m}$ of ribozymes and ${\bar y}$ of parasites grow according to 
\begin{equation}
\begin{split}
\bar{m}&=m \exp (\alpha T),\\
\bar{y}&=(n - m ) \exp (\gamma T),
\end{split}
\label{deterministic eqs}
\end{equation}
where $T$ denotes the time and $\alpha$ (resp. $\gamma$) denote the average growth rate of the ribozymes 
(resp. parasites) during this exponential growth phase.
The relevant quantity for this dynamics is the ratio of
the number of daughters of one parasite molecule 
and that of the daughters of
one ribozyme molecule:  $\Lambda=\exp ((\gamma - \alpha) T ))$. 
Note that $\Lambda >1$ since $\gamma > \alpha$. 
This exponential growth phase (maturation phase) ends, 
when the total number of templates $N=\bar{m}+\bar{y}$ 
reaches the constant value $n_{Q \beta}$ which is the same for all compartments. 
After this point, the kinetics switches to a linear one, because 
enzymes rather than templates are limiting \cite{Spiegelman1965}. 
Importantly, during this linear regime   
the ratio of ribozymes and parasites 
\be
\xb(n,m)=\frac{\bar{m}}{N}=\frac{m}{n \Lambda - (\Lambda -1) m}
\label{xb}
\ee
does not change.
Apart from neglecting very small fluctuations in $n_{Q\beta}$ and $n_u$, 
our assumption that $N$ is constant means that the effects of fluctuations of growth rates of both species 
and the effect of a possible dependence of $\Lambda$ on $m$ and $n$ are not considered. 
These two stochastic effects have been modeled in detail in the Supplemental Materials \cite{SM}. In the end, 
we find that they do not alter significantly the predictions of the present deterministic model
for the conditions of the experiment.

Two types of parasites can appear: {\it hard parasites} are formed when the replicase overlooks or 
skips a large part of the sequence of the ribozyme during replication. The resulting polymers are significantly 
shorter than that of the ribozyme and will therefore replicate much faster. 
Based on the experiments of \cite{Matsumura2016}, 
we estimated $\Lambda$ to be in a range from $10$ to about $470$, as 
explained in the Supplemental Materials \cite{SM}. 
In contrast, if the replicase makes errors but keeps overall the length of the polymers unchanged, then  %
the replication time is essentially unaffected. In that case, one speaks of {\it soft parasites}, and 
the corresponding $\Lambda$ is close to unity.
It is important to appreciate that the distinction between {\it hard} or {\it soft parasites} is not only 
a matter of replication rates, 
because $\Lambda$ also contains the time $T$, so depending on both parameters, 
parasites could be classified as either {\it hard} or {\it soft}.

The compartments are then selected according to a selection
function $f(\xb) \ge 0$. 
A specific form which is compatible with \cite{Matsumura2016} 
is the sigmoid function  
\be
\label{selection_fct}
f(\xb)=0.5 \left(1+\tanh \left( \frac{\xb-x_{th}}{x_w} \right) \right),
\ee
with $x_{th}=0.25$ and $x_w=0.1$.
Note that this function takes a small but non-zero
value for $\xb=0$, namely $0.5 ( 1 - \tanh (x_{th}/x_{w}) )= 0.0067$, which
represents the fitness of a pure parasite compartment. This is in contrast with the linear selection function 
chosen in a recent study of a similar system~\cite{Zadorin2017}. 

After the selection phase, the resulting products are pooled and the process is restarted with newly formed compartments. We
wish to evaluate the steady-state ratio $x$ of ribozymes when many rounds of the process have
taken place. 
The probability distribution of the initial condition $(n, m)$ is given by
\be
\label{def_proba}
P_\lambda (n,x, m) = {\rm Poisson}(\lambda, n) B_m(n, x),
\ee
where $B_m(n,x)$ is the Binomial distribution for $m\in\{0,\ldots,n\}$ of parameter $x \in [0, 1]$. The average of $\xb$ after the selection step is given by
\be
\label{Recursion}
x'(\lambda,x)=\frac{\sum_{n,m} \xb(n,m) f(\xb(n,m)) P_\lambda(n,x,m)}{\sum_{n,m} f(\xb(n,m)) P_\lambda(n,x,m)}.
\ee

The steady-state value of $x$ is the stable solution of
\be
x=x'(\lambda,x). 
\ee
It is easier to evaluate
$\Delta x=x'(\lambda,x) -x$ as a function of $\lambda$. The steady-state value corresponds to the line $\Delta x = 0$
separating negative values above from positive values below as shown in Fig.~\ref{fig:L2}.

\begin{figure}[htb]
\begin{center}
\includegraphics[scale=0.5]{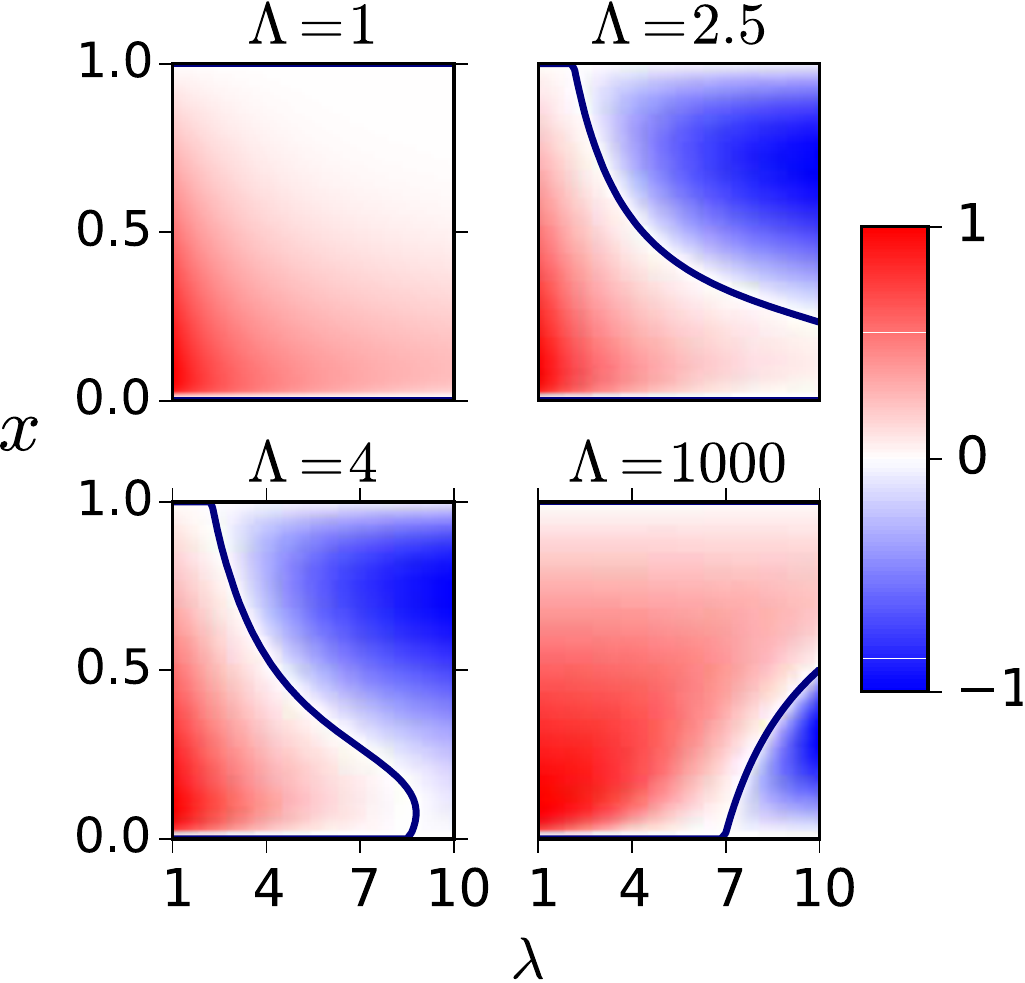}
\end{center}
\caption{Contour plots of $\Delta x$ for four values of $\Lambda=1,2.5,4$ and $1000$ 
in the plane ($x,\lambda$), with red (resp. blue) regions 
corresponding to $\Delta x > 0$ (resp. $\Delta x < 0$).} 
\label{fig:L2}
\end{figure}

We construct a phase diagram in the ($\lambda,\Lambda$) plane, by numerically evaluating the bounds of stability of the fixed point $x=0$ from the condition:
\bea
\left. \frac{\partial x' }{\partial x} \right|_{x=0} &=& 1, 
\label{fixpoint}
\eea
and similarly for the other fixed point $x=1$.
The resulting phase diagram,  as shown in Fig.~\ref{fig:phase diag}, shows four distinct phases.
In the orange (resp.\ light blue P region) region R, the only stable fixed point is $x=1$ (resp.\ $x=0$). 
In the green region, $x=0$ and $x=1$ are both stable fixed points. The system converges towards one fixed point or the other  
depending on the initial condition: for this reason we call this region B for bistable.
In the violet region, $x=0$ and $x=1$ are both unstable fixed points, but there exists a third stable fixed point 
$x^*$ with $0<x^*<1$. We call this a coexistence region (C). All of these phases can be seen in Fig.~\ref{fig:L2}. 
In the Supplemental Materials \cite{SM}, 
we discuss other aspects of the phase behavior which are not captured by this treatment. We also
show there that many features of this phase diagram remain if a linear selection function is used instead of 
Eq.~\eqref{selection_fct}

\begin{figure}[htb]
\begin{center}
\includegraphics[width=\columnwidth]{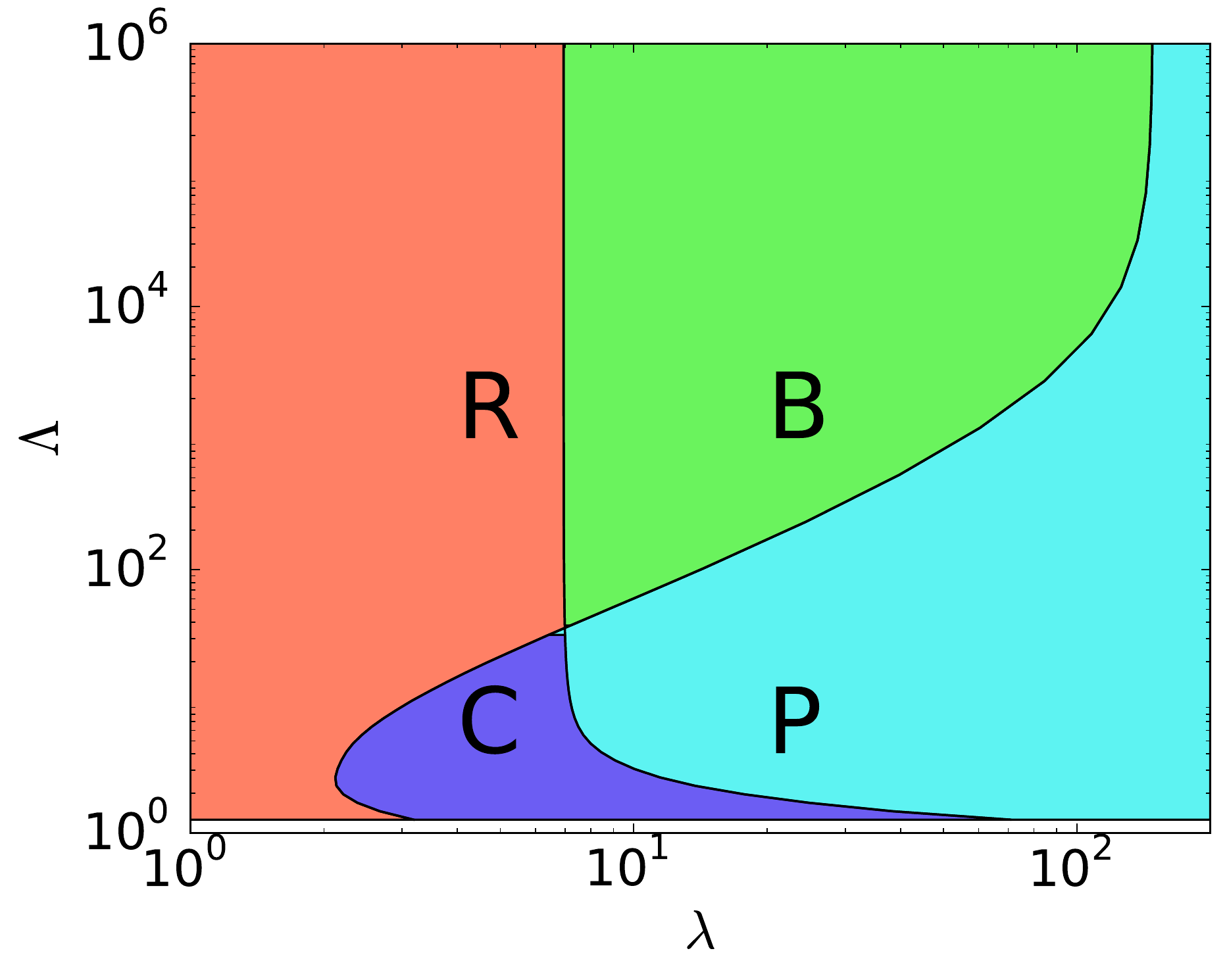}
\end{center}
\caption{Phase diagram of the transient compartmentalization dynamics with the selection function of Eq.~(\ref{selection_fct}) in
 the ($\lambda,\Lambda$) plane. The phases are: R: pure Ribozyme, B: Bistable, C: Coexistence, P: pure Parasite.}
\label{fig:phase diag}
\end{figure}

It is interesting to analyze separately some specific limits for which the 
asymptotes of the phase diagram can be computed exactly. Let us consider  
\begin{itemize}
 \item $\lambda \gg 1$: bulk behavior
 \item $\Lambda \gg 1$: {\it hard} parasites
 \item $\Lambda$ close to $1$: {\it soft} parasites
\end{itemize}

For large $\lambda$, we can neglect the fluctuations of $n$, i.e.\ the total number of replicating molecules 
(ribozymes plus parasites) 
in the seeded compartment. Indeed, $n$ is Poisson distributed with parameter $\lambda$, therefore 
Var$(n)/\lambda^2 = 1/\lambda \ll 1$. 
For large $\lambda$, $\Lambda$ close to $1$ and $x$ close to 1 (resp.\ 0), the most abundant 
compartments verify $m=n$ or $m=n-1$ (resp.\ $m=0$ or $m=1$). By considering only these compartments 
in the recursion relation \cite{SM}, one finds that the condition of stability of the fixed point $x=0$ leads to
\be
\Lambda=1 +  \frac{f'(0)}{f(0) \lambda} + O\Big(\frac{1}{\lambda^{2}} \Big), \label{HA}
\ee
for an arbitrary selection function and $\Lambda \simeq 1+ 19.86/\lambda$ for the selection function of Eq.~(\ref{selection_fct}).
This equation indeed characterizes the separation between the parasite and coexistence regime at large $\lambda$ 
in Fig.~\ref{fig:phase diag}. A similar equation is found for the fixed point at $x=1$ 
\be
\Lambda=1 +  \frac{f'(1)}{f(1) \lambda} + O\Big(\frac{1}{\lambda^{2}} \Big), \label{LA}
\ee
yielding $\Lambda \simeq 1 + 6.12\, 10^{-6}/\lambda$ for this selection function for the separation between
ribozyme and coexistence regions. 
For $\Lambda$ close enough to 1, we have a ribozyme phase. 
The asymptotes given by \eqref{HA} and \eqref{LA} border the coexistence region in Fig. \ref{fig:phase diag}. This supports the observation that {\it soft} parasites can coexist with ribozymes.

Let us now study the {\it hard} parasite limit, namely $\Lambda\gg 1$, and finite $\lambda$. 
In this regime, we only need to consider three types of compartments: compartments made of pure ribozymes, such that $m=n \neq 0$, compartments containing parasites, and empty compartments, i.e.\ such that $n=0$.
One can introduce three inoculation probabilities for these cases $p_{ribo}, p_{para}$, and $p_{zero}$. 
Using Eq.~(\ref{def_proba}), one finds
\begin{equation}
\begin{split}
p_{ribo} &= \sum_{n=1}^{\infty} \frac{x^{n} \lambda^{n}}{n!} e^{-\lambda}=( e^{\lambda x} -1) e^{-\lambda}, \\
p_{zero} &= e^{-\lambda}, \\
p_{para} &= 1-p_{ribo}-p_{zero} = 1 -e^{\lambda (x-1)}.
\end{split}
\end{equation}

Assuming that in compartments containing parasites they will overwhelm the ribozymes, and inserting these values in \eqref{Recursion}, we find
\be
x'=  \frac{p_{ribo} f(1)}{ p_{ribo} f(1) + p_{para} f(0)}.
\ee
Evaluating the fixed-point stability of $x=1$  
using \eqref{fixpoint}, we find that the boundary value of $\lambda$ satisfies
\be
\lambda f(0) e^{\lambda} = ( e^{\lambda}-1 ) f(1),
\label{vertical1}
\ee
for an arbitrary selection function.  A similar calculation at the fixed point $x=0$ leads to the other vertical separation line given by
\be
\lambda f(1)=(e^{\lambda}-1)f(0).
\label{vertical2}
\ee
The solution of Eq.~\eqref{vertical1} (resp. Eq.~\eqref{vertical2}) is $\lambda \simeq 149.41$ (resp. $\lambda \simeq 6.95$) 
which compare well with the vertical separation lines in Fig.~\ref{fig:phase diag}. 

\begin{figure}[htb]
\begin{center}
\includegraphics[scale=0.5]{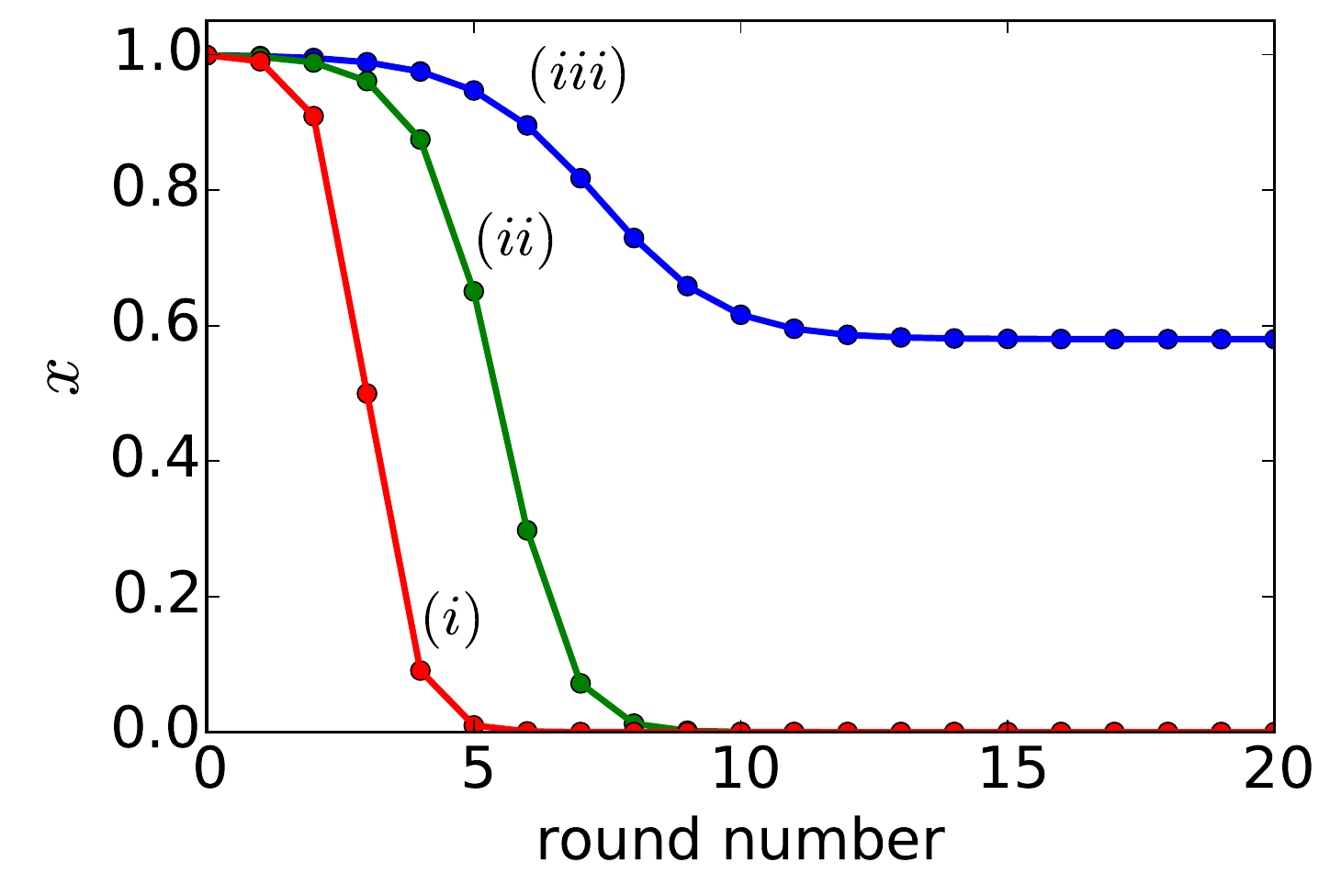}
\end{center}
\caption{Evolution of the average ribozyme fraction $x$ as function of the number of rounds for the three protocols, 
namely (i) No compartments (bulk behavior), (ii) compartments with no selection, (iii) compartments with selection. We choose  $\lambda=5$ and $\Lambda=10$, corresponding to the coexistence region of Fig.~\ref{fig:phase diag}.}
\label{fig:comparison}
\end{figure}

In ref.~\cite{Matsumura2016} a comparison was made
of the system behavior  as a function of the number of selection rounds in three possible protocols: 
(i) No compartments (bulk behavior), (ii) compartments with no selection, (iii) compartments with selection.
Such a comparison based on our theoretical model is shown in Fig.~\ref{fig:comparison} for parameter values corresponding 
to the coexistence region of Fig.~\ref{fig:phase diag}. 
As expected, the fraction of ribozymes decreases towards zero rapidly 
in case (i), and somewhat less quickly in case (ii). Only in case (iii) is it possible to maintain  
a non-zero ribozyme fraction on long times.
It is indeed observed that the ribozyme fraction eventually vanishes for protocols (i) and (ii) in the experiment of Ref.~\cite{Matsumura2016}. In 
case (iii), a decrease of the ribozyme fraction is observed. The last two points of figure~2C (top panel) in this reference are an indication that the system may eventually reach ribozyme-parasite coexistence in this regime.

\begin{figure}[htb]
\begin{center}
\includegraphics[scale=0.5]{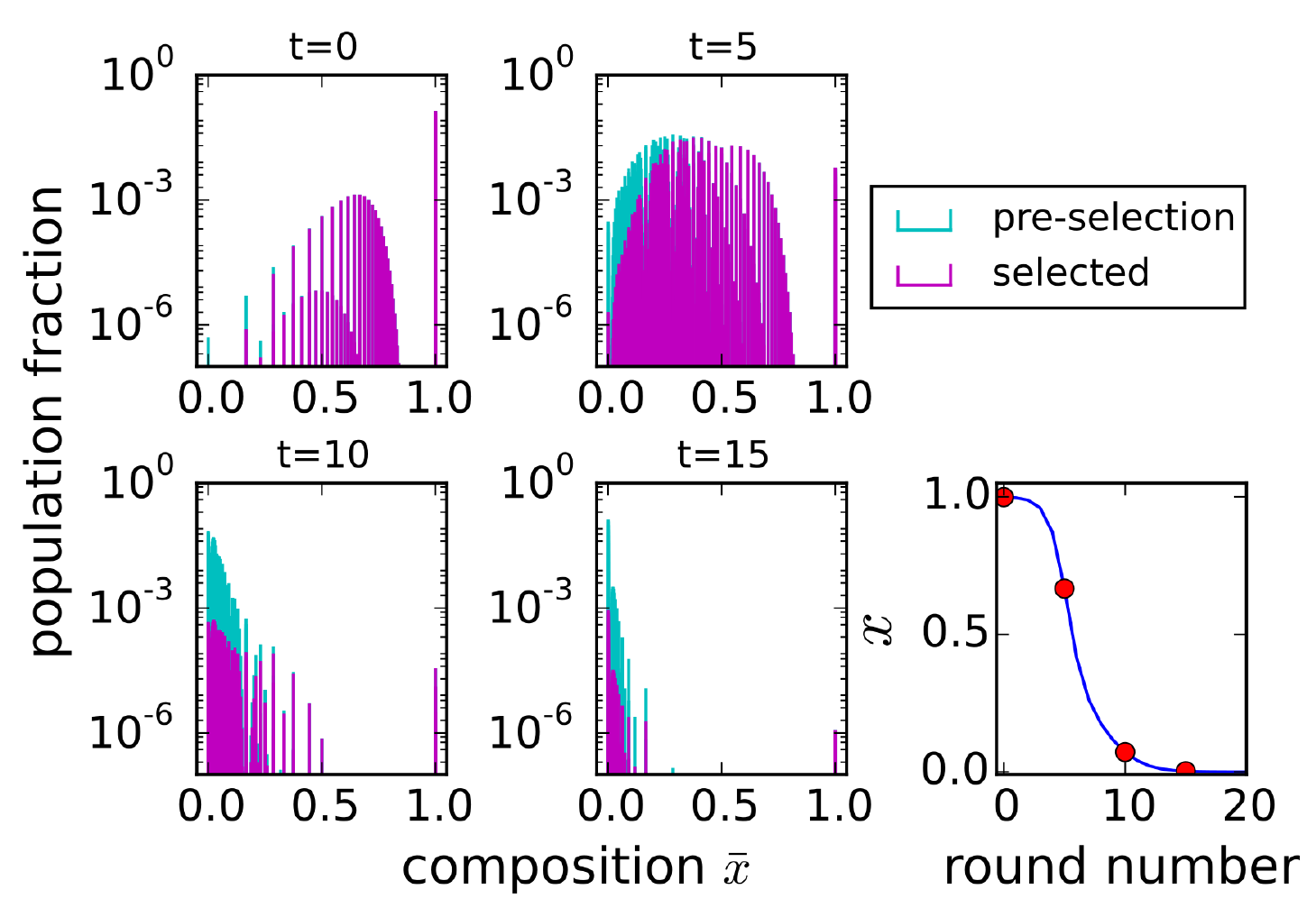}
\end{center}
\caption{Evolution of the distributions of ribozyme fraction $\xb(n,m)$ before and after selection at different times. 
The chosen times are shown as red circles in the lower right panel, which represents the evolution of the average fraction $x$
as a function of the number of selection rounds.}
\label{fig:totfig}
\end{figure}

In figure~\ref{fig:totfig} we show the behavior of the 
distribution of the ribozyme fraction after the growth phase, i.e.\ $\xb(n,m)$
(defined in Eq.~(\ref{xb})) as a function of round number.
 The parameters are $\Lambda=5$ and $\lambda=10$, corresponding to 
the parasite region, where the final state of the system is $x=0$, and the initial condition is~$x=0.999$. 
Note that the distribution of $\xb(n,m)$ is discrete, since many values are not accessible in the allowed range of $n$ and $m$.
At $t=0$, it exhibits a sharp peak near $\xb=1$ coexisting with a broad peak at small values of $\xb$. 
As time proceeds, the weight of the distribution shifts to the peak at small values of $\xb$, since in this case selection 
is not sufficiently strong 
to favor the peak near $\xb=1$ and parasites eventually take over. 

In conclusion, we captured the behavior of transient compartmentalization with a model containing only two parameters, which   
remarkably suffices to capture the main features of the transient compartimentalization experiment ~\cite{Matsumura2016}.  
The model predictions are summarized in a phase diagram, which has been derived for an arbitrary selection 
function.

Given its basic ingredients, the competition between a host and its parasite, 
and the diversity generated by small size compartments, 
which is required for selection to be efficient \cite{Fisher1930}, 
the model has broad applicability. It could for instance be relevant  
for phage-bacteria ecology problems, which typically experience a similar life cycle of transient replication in cellular compartments during infection \cite{Sneppen2014}. 

Our work also clarifies that group selection 
is able to purge the parasites even when compartments are transient. If such dynamics of compartments is applicable to 
protocells \cite{Oparin1952}, the mechanism discussed here  
 could represent an important element in scenarios on the origins of life.

A.B. was supported by the Agence Nationale de la Recherche 
(ANR-10-IDEX-0001-02, IRIS OCAV).
L.P. acknowledges support from a chair of the Labex CelTisPhysBio (ANR-10-LBX-0038).
He would like to thank ESPCI and its director, J.-F. Joanny, for a most pleasant hospitality.

\begin{widetext}

\vskip 0.8 cm

\centerline{\Large \sf Supplementary Material}
\vskip 0.3 cm
Here we provide more details on (i) the determination of the parameter $\Lambda$ from experimental data and its intercompartment variation, 
(ii) the role of growth rate fluctations, (iii)
the impact of mutations in the limit $\Lambda \to \infty$, (iv) details on the derivation of the asymptotes, 
for $\lambda\to\infty$ and~$\Lambda\simeq 1$, (v) further aspects of the phase diagram, 
and (vi) a comparison of the phase diagram for linear and non-linear selection functions. 

\section{Exponential growth and the value of $\Lambda$}
The following table contains experimental values measured in Ref.~\cite{Matsumura2016} for the ribozyme and three different parasites. The nucleotide
length, doubling time ($T_d$), and relative replication rate ($r$), are reported, from which we infer 
$\Lambda$ in the final column. The doubling time $T_d$ for the ribozyme 
is related to the growth rate $\alpha$ introduced in the main text by $T_d=\ln(2)/\alpha$, and similarly
the doubling times of the parasites is related to the $\gamma$ by $T_d=\ln(2)/\gamma$.

\begin{center}
\begin{tabular}{|c|c|c|c|c|}
  \hline
  Type  & Length (nt) 2 & $T_{d} (s)$ &  Relative $r$ & $\Lambda$ \\
  \hline
  Ribozyme & 362 & 25.0 &  1.00 & 1\\
  Parasite 1 & 245 & 20.7 &  1.21 & 13\\
  Parasite 2 & 223 & 17.1 &  1.46 & 107\\
  Parasite 3 & 129 & 14.6 &  1.71 & 473 \\
  \hline
\end{tabular}
\end{center}

In the experiment, a typical compartment contains $\lambda$ RNA molecules that can be ribozyme or parasite,
$2.6 \cdot 10^{6}$ molecules of Q$\beta$ replicase, and $1.0 \cdot 10^{10}$ molecules of each NTP. 
Replication takes place by complexation of RNA with Q$\beta$ replicase, which uses NTPs 
to make a complementary copy. This copy is then itself replicated to reproduce the original. 
There is a large amount of nucleotides, so that 
exponential growth of the target RNA proceeds until $N \approx n_{Q \beta}$. 
Starting from a single molecule, it takes $n_{D}=\log_{2} n_{Q \beta} =21.4$ doubling times to 
reach this regime. In a parasite-ribozyme mixture, we can estimate $\Lambda$ using the relative $r$:
\be
\Lambda=\frac{2^{n_D}}{2^{n_D / r }}=2^{n_D (1-\frac{1}{r})}.
\label{Lambda}
\ee

Another important assumption of our model, is that we neglect a possible dependence of $\Lambda$ on $n$ and $m$.
In order to test this assumption we have estimated the fluctuations of $\Lambda$ in the following way.
We recall that the total number of RNA at the end of the exponential phase is the constant $n_{Q\beta}$ given above, thus
$N=(n-m)  2^{n_D} + m 2^{n_D / r }=n_{Q\beta}$.
We first solve for $n_D$ in this equation for a given $n$ and $m$ and then we use this result into Eq. (\ref{Lambda}) to obtain $\Lambda$ for a given $n$ and $m$.
We show in figure \ref{fig:LL} a typical plot of the values taken by this function $\Lambda(n,m)$ for 
a particular choice of $n$ and $m$, together with the probability distribution $P_\lambda(n,x,m)$
defined in Eq. (4) of the main text. In general $\Lambda(n,m)$ is close to a constant for soft parasites ($r=1.2$), and 
is less constant for hard parasites ($r=1.7$). Even in the later case however, 
$\Lambda$ hardly varies in the range of $n,m$ values where the probability distribution 
takes significant weight.
We conclude that the assumption of neglecting a possible dependence of $\Lambda$ on $n$ and $m$ has only 
a minor effect on our results.
\begin{figure}[!htb]
\begin{center}
\includegraphics[scale=0.4]{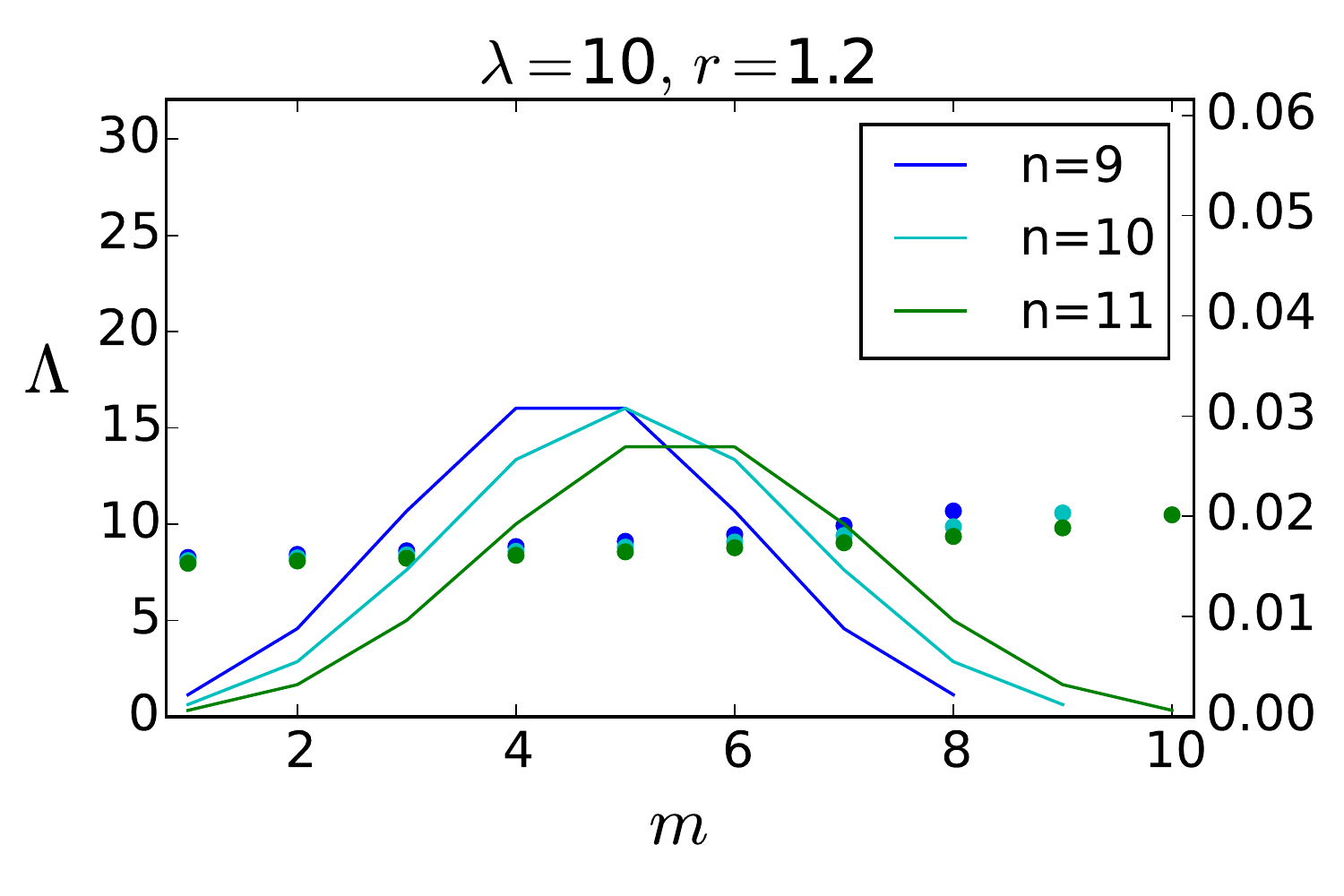}
\includegraphics[scale=0.4]{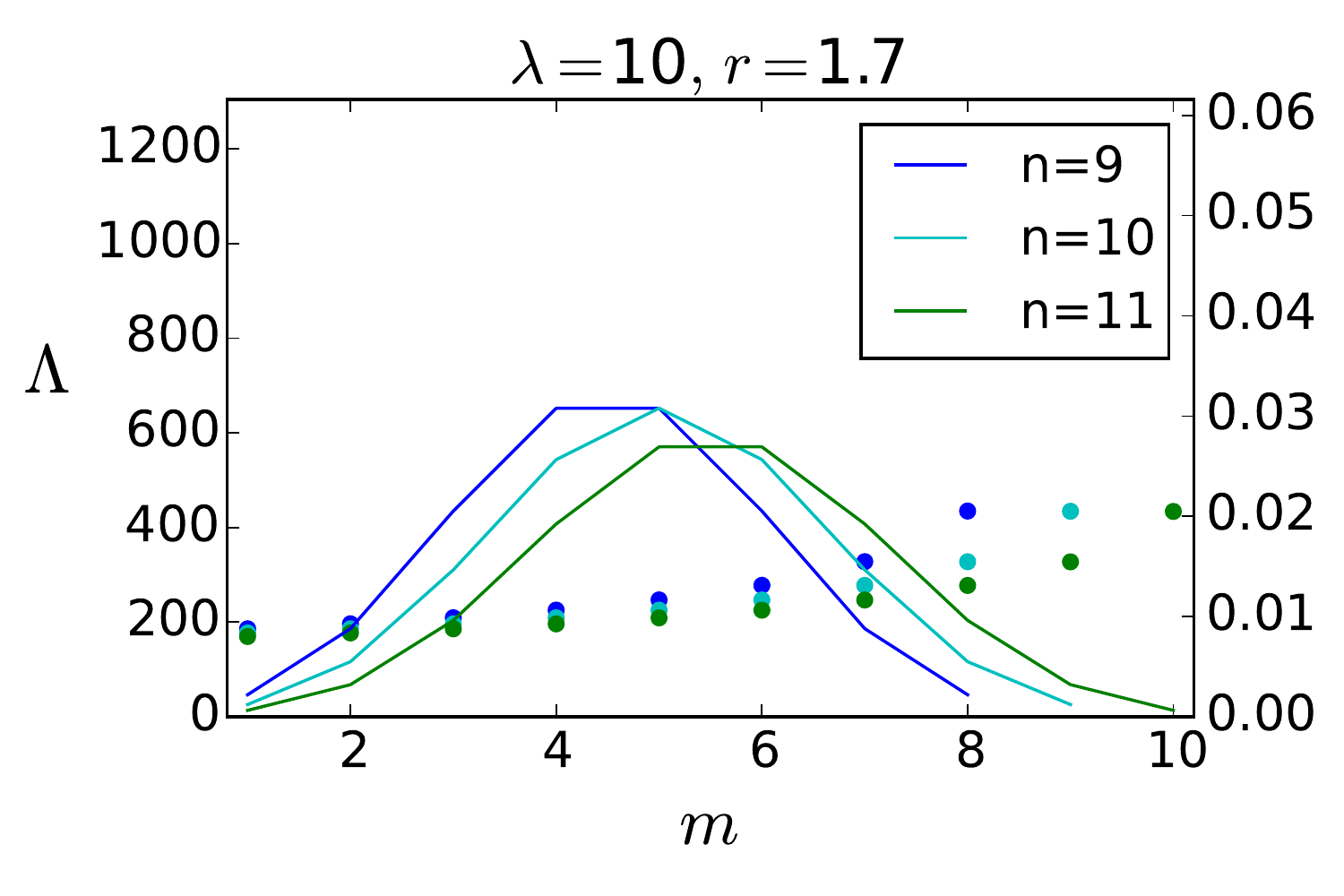}
\end{center}
\caption{Plots for $\Lambda$ (colored dots) as function of the parameters $(n,m)$ characterizing the initial composition 
of the compartments relative growth rates $r$ and $\lambda=10$, together with the probability distribution $P_\lambda(n,x,m)$ (solid lines) for $x=0.5$.
\label{fig:LL}}
\end{figure}

\section{Growth rates fluctuations}
The model presented in the main text is purely deterministic, but 
fluctuations are still present due to the initial condition. 
Since growth rates typically depend on the initial condition, they will fluctuate 
when sampling the initial condition. 
While these fluctuations are already present in a deterministic model, 
effects associated to other fluctuations 
in the growth rates, which may occur at time scales smaller than the duration of the growth phase
require a stochastic approach. 
Replication is intrinsically stochastic and therefore this additional source of 
fluctuations in grow rates could be present.
In order to estimate such an effect, we have implemented below a stochastic version of our
model. 

A stochastic component in the growth phase of the ribozymes 
and parasites can be included using a discrete Langevin approach. 
In such an approach, Eq. (1) is modified to become:
\be
\ln \frac{\bar{m}}{m}=\alpha T + \xi_1,
\ee
where the first term on the right hand side accounts for the deterministic contribution
 we had before and $\xi_1$ is a Gaussian random variable of mean zero and 
variance $\sigma_1$. 
Similarly for the parasites:
\be
\ln \frac{\bar{y}}{y}=\gamma T + \xi_2,
\ee
where $\xi_2$ is another similar noise controlling the growth of the parasites.

The noise which has been introduced here could describe  
either fluctuations of growth rate $\alpha$ or of the duration of replication. Note also that we still define $\Lambda$ with respect to the mean time and growth rates as 
$\Lambda=\exp ((\gamma - \alpha) T ))$. Since $N$ is fundamentally fixed by 
the number of enzymes $n_{Q\beta}$ in this problem,  
we still assume that $N$ is fixed in this stochastic model.
Then this condition $N=\bar{m}+ \bar{y}$ leads to a constraint between 
the noise $\xi_1$ and $\xi_2$, which means that these two noises must be correlated.
Then, Eq (2) is modified as
\be
\xb(n,m,\eta)=\frac{\bar{m}}{N}=\frac{m}{m + (n-m) \Lambda \exp{(\eta)}},
\label{xb}
\ee
where we have introduced the random variable $\eta=\xi_2 -\xi_1$. Let us introduce the variance of $\eta$ which we call $\sigma^2$. This is the main parameter controlling the growth noise. 

Eq. (5) is now modified as follows
\be
\label{Recursion}
x'(\lambda,x)=\frac{\int d\eta g(\eta) \sum_{n,m} \xb(n,m,\eta) f(\xb(n,m,\eta)) P_\lambda(n,x,m)}{\int d\eta g(\eta) \sum_{n,m} f(\xb(n,m,\eta)) P_\lambda(n,x,m)},
\ee
where $g(\eta)$ is the Gaussian distribution of mean zero and variance $\sigma^2$.
In order to evaluate the correction due to the noise $\eta$, we expand the integrand in the numerator 
and denominator with respect to $\eta$ and we perform the Gaussian integrals.
The result is a modified recursion relation which contains a correction term proportional to $\sigma^2$.
The explicit expression of this correction is lengthy and not given here, it was evaluated numerically.

This Taylor expansion is justified if the amplitude of the noise $\sigma^2$
is sufficiently small. 
In order to assess this, we investigate the various sources of noises in this problem.
The noise could be due to the arrival times of $Q\beta$ or
from the replication process itself.
For the first source of noise, the time scale to form an $Q \beta$ - RNA complex 
due to diffusion can be evaluated as  
 $ t_D \approx (D_{RNA} \ c_{Q\beta}^{\frac{2}{3}})^{-1}$, 
where $D_{RNA}$ is the diffusion constant for an RNA strand (length $\approx 300$) and $c_{Q \beta}$ the concentration of $Q \beta$ replicase. 
We found this timescale to be over $2 \ 10^4$ times smaller than replication times (15-25s, see SM 1). 
Due to this large difference in timescales, the noise in this problem should primarily be caused 
by the replication rather than by the binding of a $Q\beta$ to an RNA strand.
Let us now look at the noise due to replication.
Once a $Q \beta$ enzyme is bound to a single RNA molecule, the total replication time $\tau$
can be written as a sum of the dwell times of all the nucleotides to be added to the template and 
which are themselves exponentially distributed. When the binding rates are identical, the resulting
distribution of $\tau$ is a gamma distribution with a coefficient of variation $1/\sqrt{n}$, in terms
of $n$ the number of nucleotides as shown in D. Floyd et al. (2010). 
Since our replicating molecules are long, typically between $n\simeq 150$ to $300$, 
this coefficient of variation is quite small.
This coefficient of variation is expected to increase due to the number of doubling times 
in the replicating phase, which is typically of the order of 20. 
In the end although we can not provide an accurate estimate
of the noise of $\eta$, all these factors suggest a small noise amplitude for $\sigma^2$.

In the worst possible case, we would have $\sigma^2=1$ which is the case shown in the two figures below. 
The red contour plots of $\Delta x$ corresponding to that of Fig 1 of the manuscript, the blue ones
 correspond to the prediction of a stochastic version of the same model including the correction 
due to $\sigma^2$.
We only show the plots for $\Lambda=4$ (left) and $\Lambda=2.5$ (right), because we find that for 
high values of $\Lambda$ the noise has only a negligible effect even in this worst case scenario,
which is reasonable.
\begin{figure}[!htb]
\begin{center}
\includegraphics[scale=0.3]{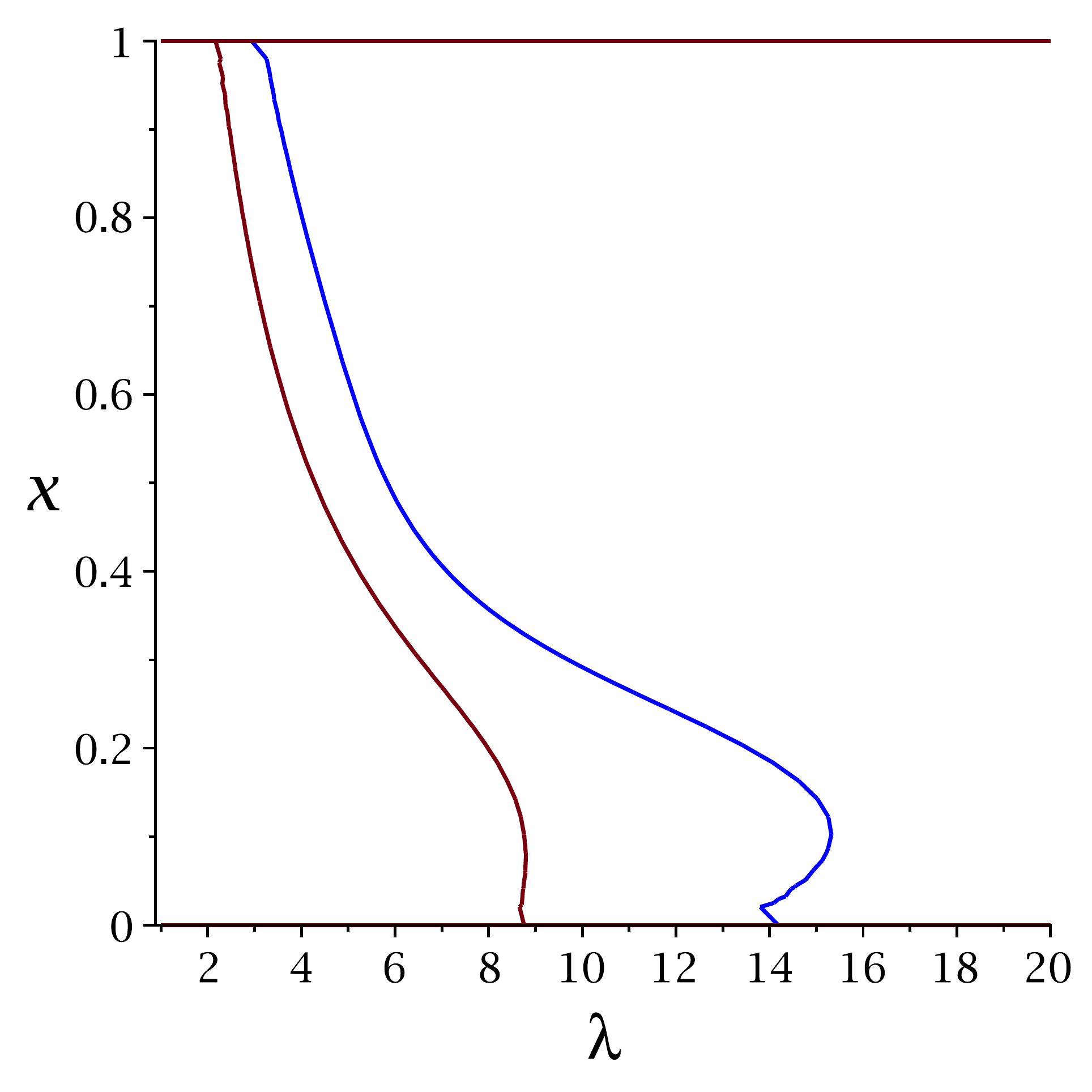}
\includegraphics[scale=0.3]{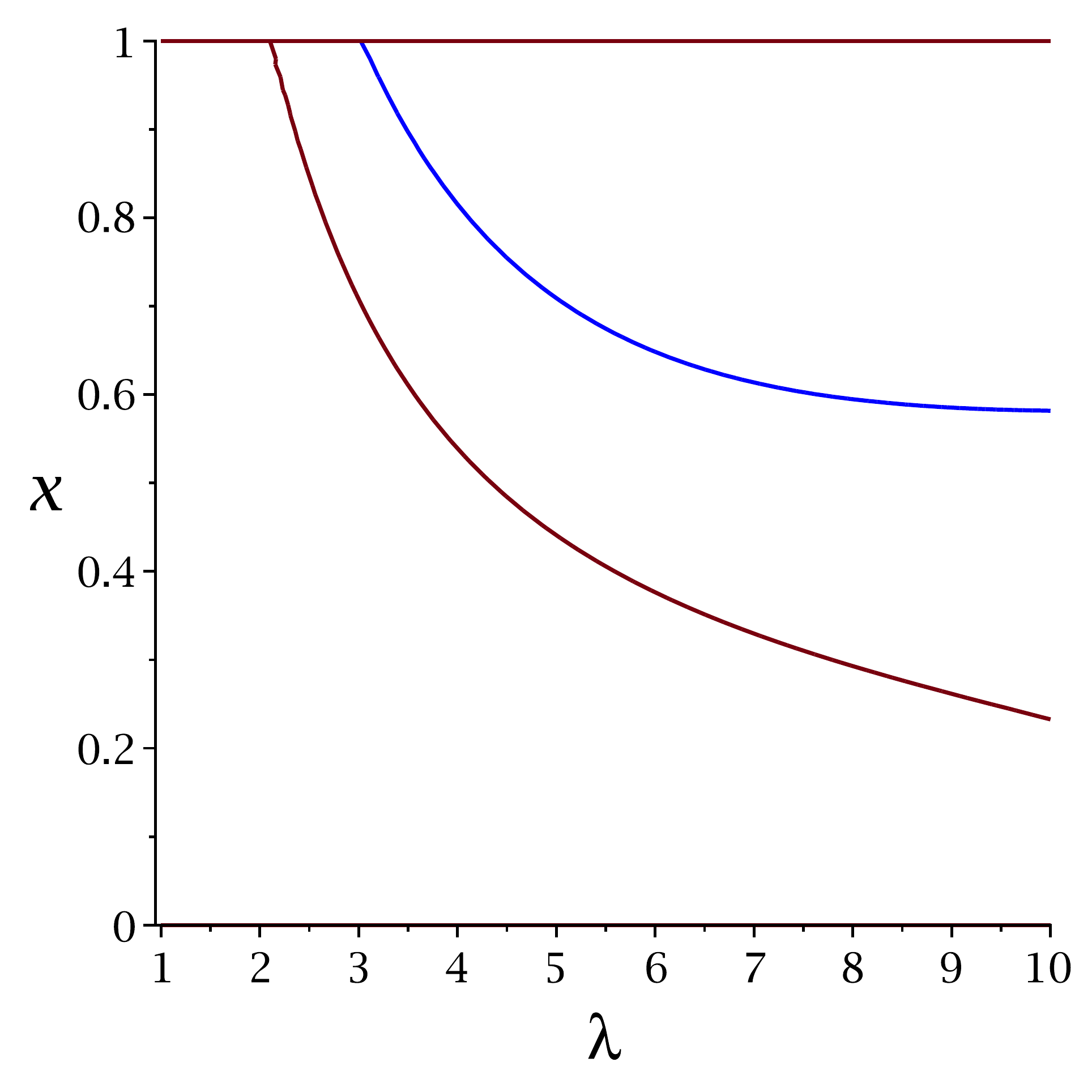}
\end{center}
\caption{Contour plots of $\Delta x$ similar to that of Fig 1 of the manuscript, with no correction 
due to noise (red solid line) and 
including corrections due to noise with $\sigma^2=1$ (blue solid line). 
The figures correspond to $\Lambda=4$ (left) and $\Lambda=2.5$ (right).}
\label{fig:1}
\end{figure}
While we see that the noise $\eta$ affects significantly the contour plots in this worst case scenario, 
when $\sigma^2=1$, the effect is quite small with a more realistic estimate of the noise
namely $\sigma^2=0.1$ as shown in figure \ref{fig:2}:
\begin{figure}[!htb]
\begin{center}
\includegraphics[scale=0.3]{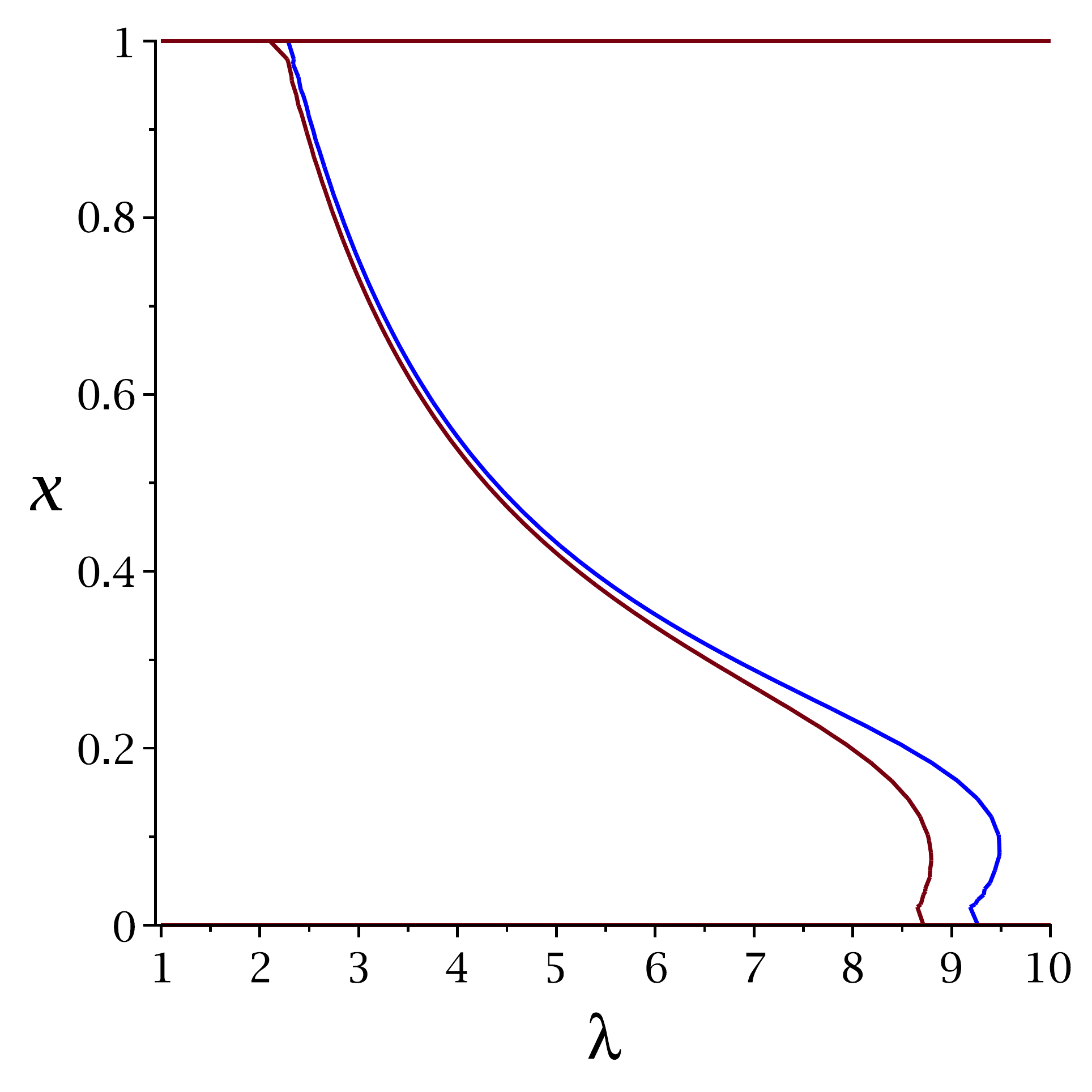}
\includegraphics[scale=0.3]{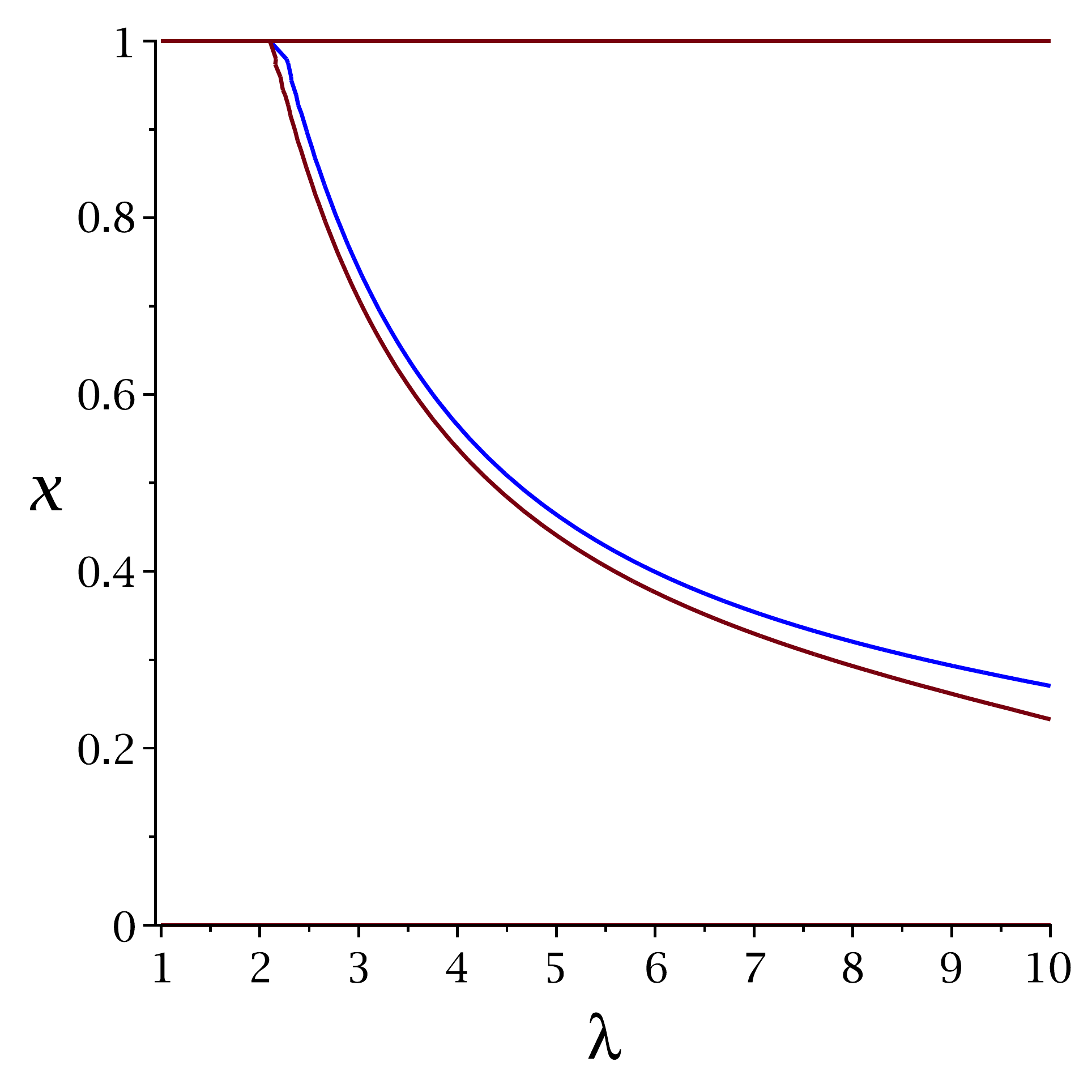}
\end{center}
\caption{Contour plots of $\Delta x$ similar to that of Fig 1 of the manuscript, with no correction 
due to noise (red solid line) and 
including corrections due to noise with $\sigma^2=0.1$ (blue solid line). 
The figures correspond to 
$\Lambda=4$ (left) and $\Lambda=2.5$ (right).}
\label{fig:2}
\end{figure}

All these results support our view that while the growth is intrinsically stochastic the deterministic 
model we have developed with a constant $N$ and 
fluctuations only in the initial conditions  
indeed capture the main features of the experiment we are interested in. 
As shown in figures $\ref{fig:1}$ and $\ref{fig:2}$, the stability of the ribozyme phase 
in the new stochastic model is enhanced with respect to the deterministic model. 
This confirms that fluctuations are essential to stabilize the ribozyme phase, and favor it whether they
 come from the initial condition as in the deterministic model or from other sources as in the stochastic developed here. 

\section{Impact of mutations in the limit $\Lambda\gg 1$}

In this section, we explain how the approach described in the main text needs to be amended in the presence of 
mutations. We focus here only on the case that $\Lambda\gg 1$, because one can expect that the effect of mutations will 
be more dramatic in this limit corresponding to {\it hard} parasites. 
If one of such parasites is present in a compartment, it invades the population very quickly, 
provided it appears during the exponential growth phase. 
As explained in the main text, to describe the limit $\Lambda\gg 1$, we can introduce three inoculation probabilities, $p_{ribo}$ 
for compartments containing only ribozymes, $p_{para}$ for compartments containing parasites and $p_{zero}$ for empty compartments:
\begin{equation}
\begin{split}
p_{ribo} &= \sum_{n=1}^{\infty} \frac{x^{n} \lambda^{n}}{n!} e^{-\lambda}=( e^{\lambda x} -1) e^{-\lambda}, \\
p_{zero} &= e^{-\lambda}, \\
p_{para} &= 1-p_{ribo}-p_{zero} = 1 -e^{\lambda (x-1)}.
\end{split}
\end{equation}

Now, let us also introduce $p_{mut}$ as the probability that a ribozyme is turned into a parasite as a result of a mutation 
during one replication event of the molecule. Then the probability that there is no mutation occurring during $n^*$ replication events is 
\be
\zeta=(1-p_{mut})^{n*}.
\ee
A typical value for this $n^*$ corresponds to what is denoted $n_D$ in the previous section, namely 
the number of replications until the end of the exponential growth regime. 

With $p_{zero}$ unchanged, the new probabilities for compartments 
containing ribozyme (resp. parasites) $p'_{ribo}$ (resp. $p'_{para}$) are simply
\bea
p'_{ribo} &=&p_{ribo} \zeta, \\
p'_{para}&=&p_{para}+p_{ribo}(1-\zeta).
\eea
Using these expressions in the recurrence relation, we obtain
\be
x' =  \frac{p'_{ribo} f(1)}{ p'_{ribo} f(1) + p'_{para} f(0)} \\
\ee

Evaluating the fixed point stability at $x=0$ then yields the equation for the asymptote
\be
\zeta \lambda  f(1) = f(0) (e^{\lambda}- 1).
\label{eq1}
\ee
Similarly, we can evaluate the fixed point stability at $x=1$, to obtain
\be
(e^{\lambda}-1) \left( 1+ \zeta \left( \frac{f(1)}{f(0)}  -1 \right) \right)^{2} = \lambda \frac{f(1)}{f(0)} e^{\lambda}.
\label{eq2}
\ee
For $p_{mut}\rightarrow 0$, $\zeta \rightarrow 1$ and we obtain the asymptotes mentioned in the text. 
For $\zeta<1$, the asymptotic values of $\lambda$ for both $x=0$ and $x=1$ become smaller.
As a result, both the ribozyme and the bistable regions shrink as one would expect. In the extreme case where $\zeta \to 0$,
both regions disappear completely since then the only solution to Eqs.~(\ref{eq1}-\ref{eq2}) corresponds to $\lambda=0$.

\section{Asymptotic behavior for $\lambda \rightarrow \infty$}
For large $\lambda$, for $\Lambda$ close to $1$ and $x$ close to 1 (resp.\ 0), the most abundant 
compartments verify $m=n$ or $m=n-1$ (resp.\ $m=0$ or $m=1$). 
As $\lambda$ is large, we can neglect fluctuations in $n$ and we can take $n=\lambda$. 
We therefore only look at the recursion for a typical compartment with $n=\lambda$, with 
a simplified notation $P_\lambda(n=\lambda,x,m)=P_\lambda(x,m)$, where 
\begin{equation}
P_{\lambda}(x,m) = B_m (\lambda,x),
\end{equation}
obtaining
\begin{equation}
x'=\frac{ f(1) P_{\lambda}(x,\lambda) + \bar{x} P_{\lambda}(x,\lambda-1) f(\bar{x}) }{ f(1) P_{\lambda}(x,\lambda) + 
P_{\lambda}(x,\lambda-1) f(\bar{x}) },
\end{equation}
where
\begin{equation}
\xb=\xb(\lambda,\lambda-1)=\frac{\lambda-1}{\lambda+\Lambda-1}\simeq 1-\frac{\Lambda}{\lambda}.
\end{equation}
We have therefore
\begin{equation}
x'=\frac{x^{\lambda}f(1)+\lambda x^{\lambda-1}(1-x)\xb f(\xb)}{x^{\lambda}f(1)+\lambda x^{\lambda-1}f(\xb)}=\frac{x f(1)+\lambda (1-x)\xb f(\xb)}{x f(1)+\lambda(1-x)f(\xb)}.
\end{equation}
Taking the derivative with respect to~$x$ we obtain
\begin{equation}
\frac{d x'}{d x}=\frac{\lambda (1-\xb)f(1)f(\xb)}{(x f(1)+\lambda(1-x)f(\xb))^{2}},
\end{equation}
which for $x=1$ yields
\begin{equation}
\left.\frac{d x'}{d x}\right|_{x=1}=\frac{\lambda(1-\xb)f(\xb)}{f(1)}.
\end{equation}
Thus the boundary defined by the equation
\begin{equation}
\left.\frac{d x'}{d x}\right|_{x=1}=1,
\end{equation}
is given by
\begin{equation}
\Lambda \simeq 1+\frac{f'(1)}{f(1)\lambda}=1+6.12\, 10^{-6}/\lambda.
\end{equation}
Evaluating the stability around the fixed point $x=0$ we obtain likewise
\begin{equation}
x'=\frac{\lambda x (1-x)^{\lambda-1}\xb f(\xb)}{(1-x)^{\lambda}f(0)+\lambda x (1-x)^{\lambda-1} f(\xb)}=\frac{\lambda x \xb f(\xb)}{(1-x)f(0)+\lambda x f(\xb) },
\end{equation}
where now $\xb$ is given by
\begin{equation}
\xb=\xb(\lambda,1)=\frac{1}{(\lambda-1)\Lambda+1}\simeq \frac{1}{\Lambda\lambda}.
\end{equation}
Evaluating the derivative of $x'(x)$ at $x=0$ we obtain
\begin{equation}
\left.\frac{d x'}{d x}\right|_{x=0}=\frac{\lambda \xb f(\xb)}{f(0)}.
\end{equation}
This gives the boundary as
\begin{equation}
\Lambda=1+\frac{f'(0)}{f(0)\lambda}=1+19.8661/\lambda.
\end{equation}

\section{Additional features of the phase diagram}

\begin{figure}
\begin{center}
\includegraphics[scale=0.5]{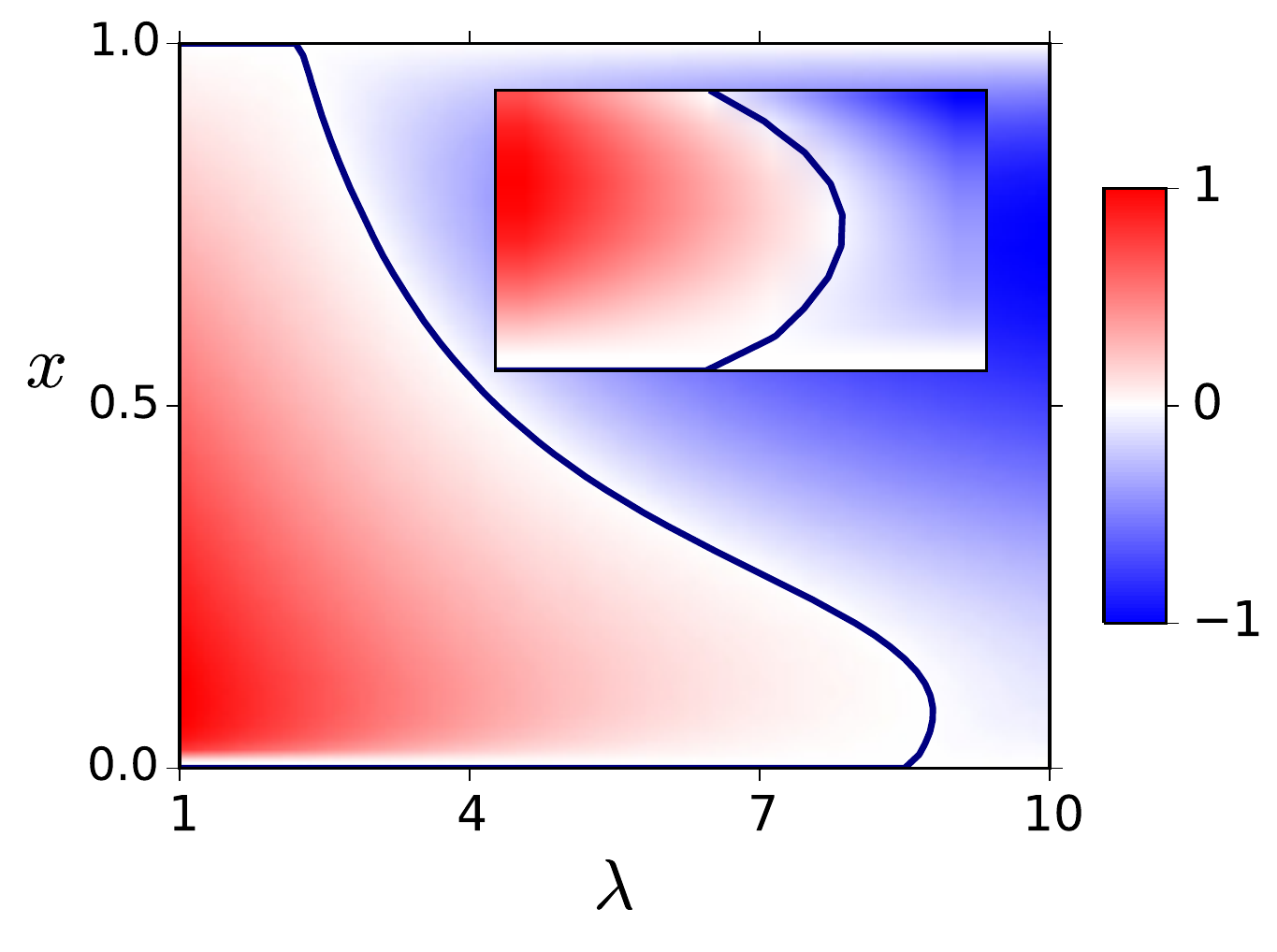}
\end{center}
\caption{Contour plots of $\Delta x$ vs.~$x$ for $\Lambda=4$ in the plane ($x,\lambda$). 
Inset shows a blow-up of the region near $\lambda=8$, which exhibits features of both the bistable and coexistence regions.}
\label{fig:L2}
\end{figure}
The construction of the phase diagram of the main text is based on the condition of stability of the two 
fixed points $x=0$ and $x=1$. 
This treatment only gives a complete picture of the phase behavior if there are at most three fixed points. 
While this is true for most pairs ($\lambda$,$\Lambda$), notice that in the special case of Fig.~\ref{fig:L2}a 
for $\Lambda=4$ and near $\lambda=8$, the curve turns back. 
In this region, there are four fixed points, with $x=0$ and $0<x^{*}<1$ being stable.
The novel aspect of the region near $\lambda=8$ and $x$ below 0.1 (shown in the inset as a blow-up), 
is that there is a bistability between points $x=0$
and $x=x^*$ whereas in the phase diagram of the main text, the bistability only concerned points $x=0$ and $x=1$. 
For $x>0.1$ and $\lambda$ between approximately 2 and 8, we have a standard coexistence phase.

\section{Comparison between linear and non-linear selection function}

In the main text, we have introduced the following selection function
\be
f(\bar{x})=0.5 \left(1+\tanh \left( \frac{\bar{x}-x_{th}}{x_w} \right) \right),\label{fsel}
\ee
with $x_{th}=0.25$ and $x_w=0.1$, which is now represented as the blue solid line in fig.~\ref{fig:select}.
Note that this selection function takes a small but non-zero
value for $x=0$, namely $0.5 ( 1 - \tanh (x_{th}/x_{w}) )= 0.0067$, which
represents the fraction of false positives in the selection process. 

\begin{figure}[htb]
\begin{center}
\includegraphics[scale=0.5]{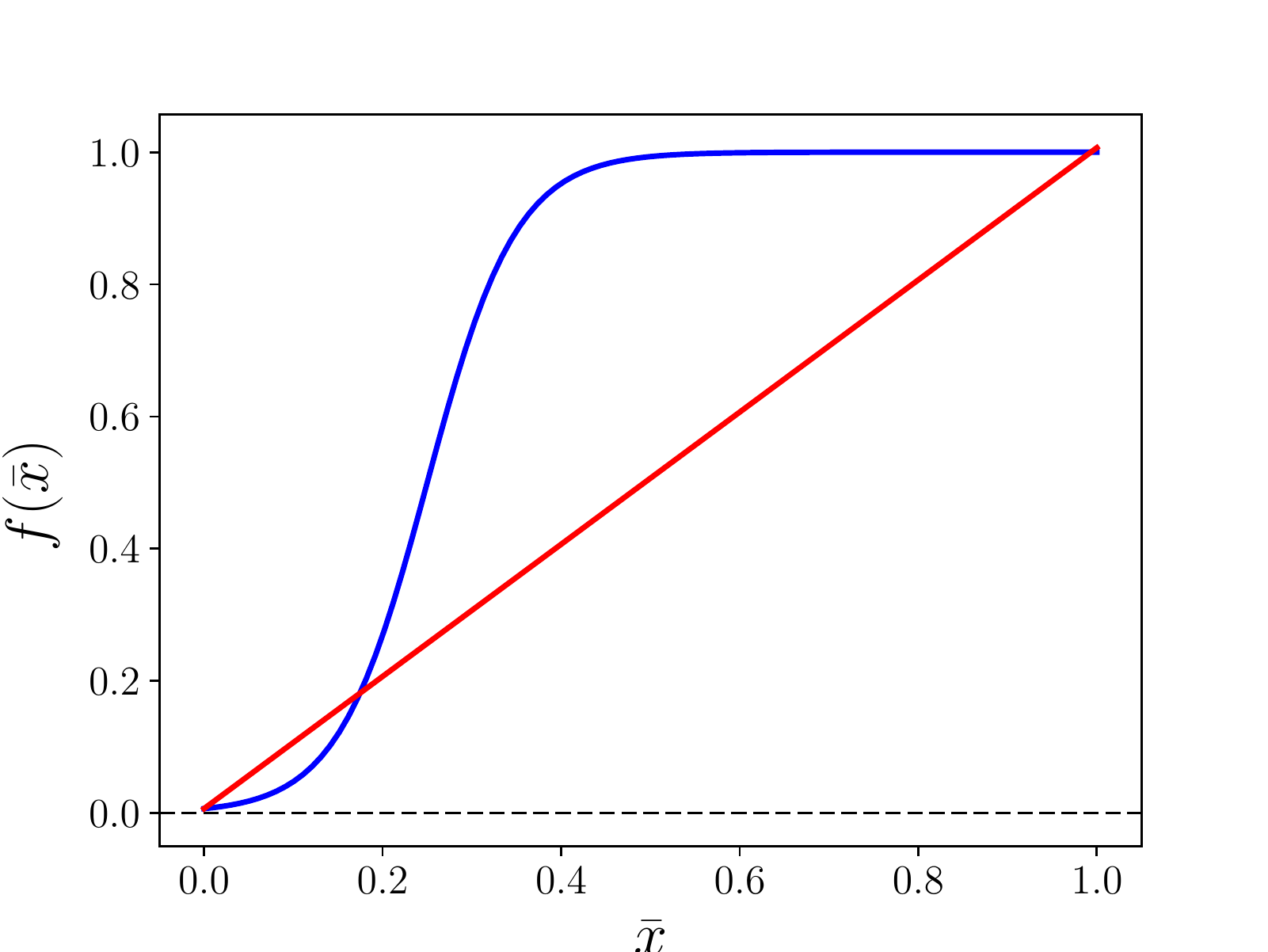}
\end{center}
\caption{Representation of the two selection functions used in this section, namely the one defined in the main text (red solid line) and 
a linear one (blue solid line), which approximately take the same values at $x=0$ and $x=1$. The horizontal dashed line represents the level 
of false positives given by $f(0)$.}
\label{fig:select}
\end{figure}

It is interesting to compare the phase diagram given in the main text with that obtained for a linear selection function, $f_{lin}$
shown as the red solid line in fig.~\ref{fig:select}. 
We choose $f_{lin}(\bar{x})=0.0067+\bar{x}$, such that we have the same approximate values for $f(0)$ and $f(1)$ as with the previous 
function defined in Eq.~\eqref{fsel}. 
Consequently, we expect to find the same vertical asymptotes at $\lambda \simeq 6.95$ and $\lambda \simeq 149$, as confirmed in 
fig.~\ref{fig:phase diag}. The equations of these vertical asymptotes are indeed only a function of $f(0)$ and $f(1)$. They are
\be
\lambda f(0) e^{\lambda} = ( e^{\lambda}-1 ) f(1),
\ee
for the boundary between the bistable and parasite regions, and 
\be
\lambda f(1)  = ( e^{\lambda}-1 ) f(0),
\ee
for the boundary between the bistable and ribozyme regions.

When comparing the two phase diagrams obtained with the non-linear and linear selection functions shown in fig.~\ref{fig:phase diag},
we observe a similar general structure except for the center of the diagram and for the two asymptotes for $\Lambda$ close to 1.
This is to be expected for the center region where none of the simple approximations hold.
Concerning the asymptotes near $\Lambda=1$, as shown in the main text they represent 
boundaries between the coexistence and parasite regions
(resp.\ coexistence and ribozyme regions) and they 
depend on the logarithmic derivative of the selection function near $x=0$ (resp.\ $x=1$).
 
\begin{figure}
\begin{center}
\includegraphics[scale=0.4]{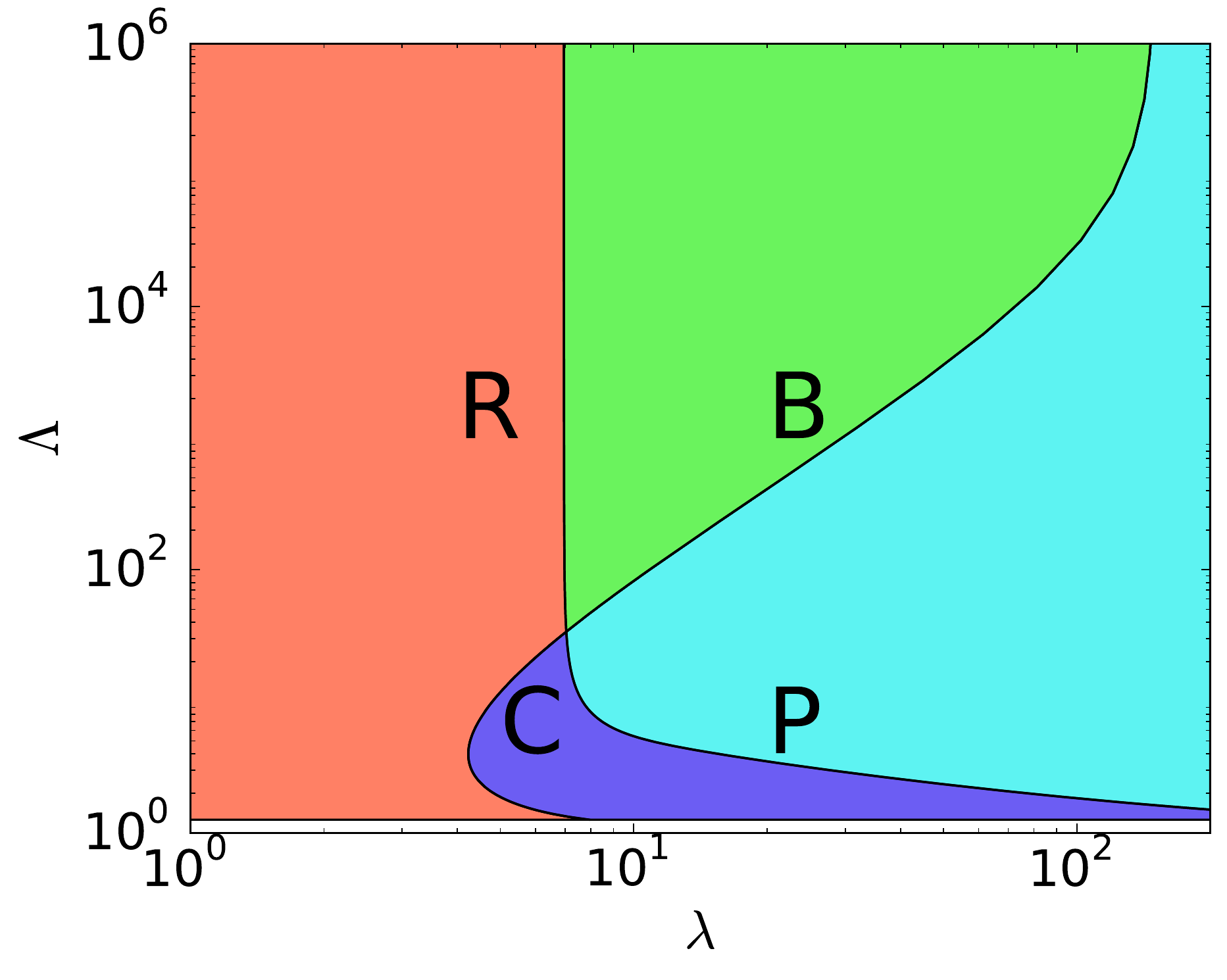}
\includegraphics[scale=0.4]{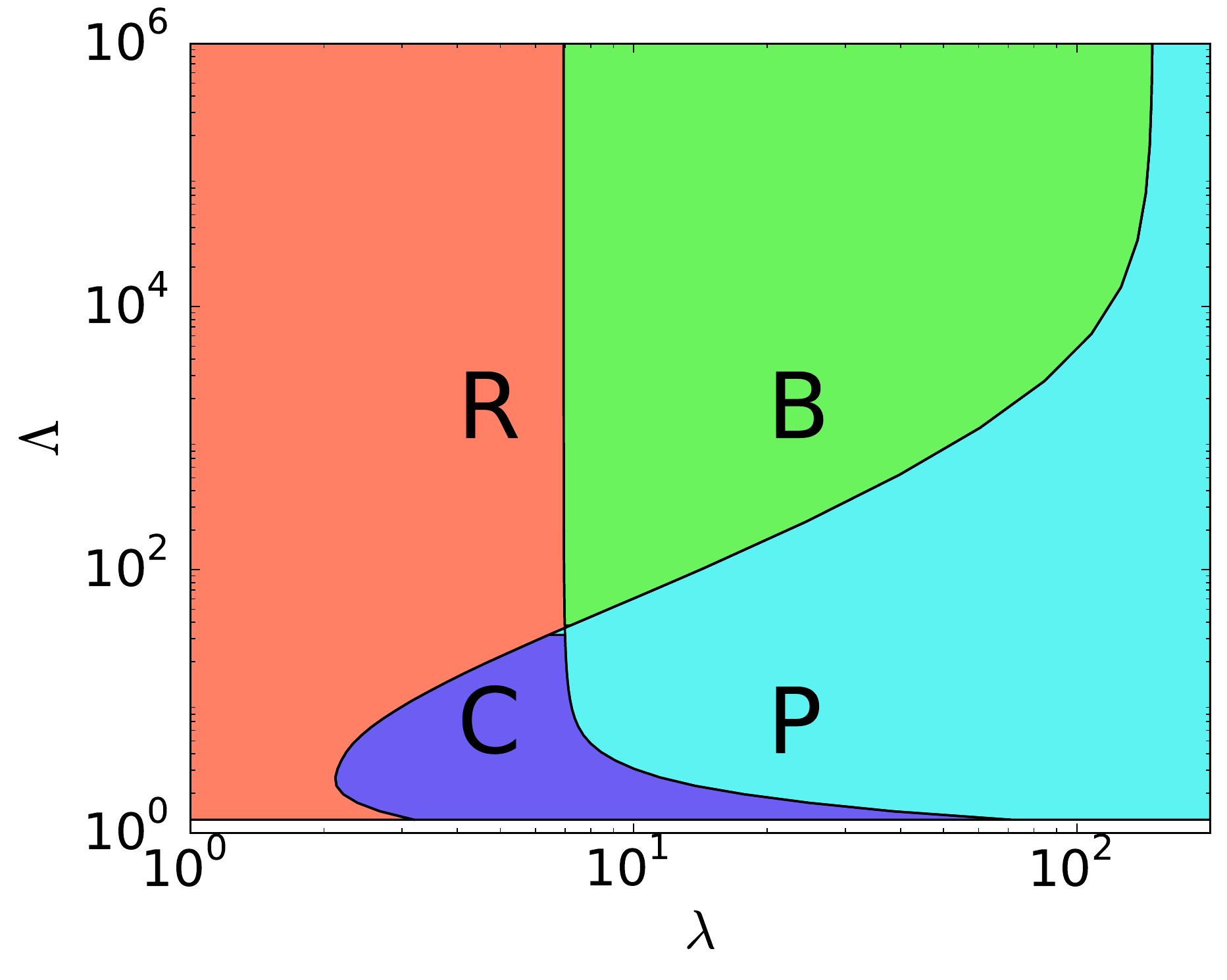}
\end{center}
\caption{Left: Phase diagram of the transient compartmentalization dynamics with the linear selection function $f(\bar{x})=0.0067+\bar{x}$ in
 the ($\lambda,\Lambda$) plane. Right: idem with the non-linear selection function represented in Fig.~\ref{fig:select}.
Phases are R: pure Ribozyme, B: Bistable, C: Coexistence and P: pure Parasite.
 }
\label{fig:phase diag}
\end{figure}
\vskip 100pt
\end{widetext}


\begin{thebibliography}{0}%
\makeatletter
\providecommand \@ifxundefined [1]{%
 \@ifx{#1\undefined}
}%
\providecommand \@ifnum [1]{%
 \ifnum #1\expandafter \@firstoftwo
 \else \expandafter \@secondoftwo
 \fi
}%
\providecommand \@ifx [1]{%
 \ifx #1\expandafter \@firstoftwo
 \else \expandafter \@secondoftwo
 \fi
}%
\providecommand \natexlab [1]{#1}%
\providecommand \enquote  [1]{``#1''}%
\providecommand \bibnamefont  [1]{#1}%
\providecommand \bibfnamefont [1]{#1}%
\providecommand \citenamefont [1]{#1}%
\providecommand \href@noop [0]{\@secondoftwo}%
\providecommand \href [0]{\begingroup \@sanitize@url \@href}%
\providecommand \@href[1]{\@@startlink{#1}\@@href}%
\providecommand \@@href[1]{\endgroup#1\@@endlink}%
\providecommand \@sanitize@url [0]{\catcode `\\12\catcode `\$12\catcode
  `\&12\catcode `\#12\catcode `\^12\catcode `\_12\catcode `\%12\relax}%
\providecommand \@@startlink[1]{}%
\providecommand \@@endlink[0]{}%
\providecommand \url  [0]{\begingroup\@sanitize@url \@url }%
\providecommand \@url [1]{\endgroup\@href {#1}{\urlprefix }}%
\providecommand \urlprefix  [0]{URL }%
\providecommand \Eprint [0]{\href }%
\providecommand \doibase [0]{http://dx.doi.org/}%
\providecommand \selectlanguage [0]{\@gobble}%
\providecommand \bibinfo  [0]{\@secondoftwo}%
\providecommand \bibfield  [0]{\@secondoftwo}%
\providecommand \translation [1]{[#1]}%
\providecommand \BibitemOpen [0]{}%
\providecommand \bibitemStop [0]{}%
\providecommand \bibitemNoStop [0]{.\EOS\space}%
\providecommand \EOS [0]{\spacefactor3000\relax}%
\providecommand \BibitemShut  [1]{\csname bibitem#1\endcsname}%
\let\auto@bib@innerbib\@empty
\end{thebibliography}%


\begin{thebibliography}{99}

\bibitem{Higgs2015} P. G. Higgs and N. Lehman, Nat. Rev. Genet. {\bf 16}, 7 (2015).

\bibitem{Eigen1971} M. Eigen, Naturwissenschaften {\bf 58}, 465 (1971).

\bibitem{Takeuchi2012} N. Takeuchi and P. Hogeweg, Physics of Life Reviews {\bf 9}, 219 (2012).

\bibitem{Levin2017} S. R. Levin and S. A. West, Proc. R. Soc. B {\bf 284}
(2017).

\bibitem{Tupper2017} A. S. Tupper and P. G. Higgs, J. Theor. Biol. {\bf 428}, 34 (2017).

\bibitem{Kim2016} Y. E. Kim and P. G. Higgs, 	PLOS Comput. Biol. {\bf 11}, 34 (2016).

\bibitem{Szathmary1987} E. Szathm\'ary and L. Demeter, J. Theor. Biol. {\bf 128},
463 (1987).

\bibitem{Smith1995} J. Maynard Smith and E. Szathm\'ary, The Major
Transitions in Evolution (Freeman, Oxford, 1995).

\bibitem{Oparin1952} A. I. Oparin, Origin of Life (Dover, 1952).

\bibitem{Zwicker2017} D. Zwicker, R. Seyboldt, C. A. Weber, A. A. Hyman,
and F. J\"ulicher, Nat. Phys. {\bf 13}, 408 (2017).

\bibitem{Brangwynne2009} C. P. Brangwynne, C. R. Eckmann, D. S. Courson,
A. Rybarska, C. Hoege, J. Gharakhani,
F. J\"ulicher, and A. A. Hyman, Science {\bf 324}, 1729
(2009).

\bibitem{hydrovent} E. V. Koonin and W. Martin, Trends Genet. {\bf 21},
647 (2005).

\bibitem{vesicles} P. L. Luisi, P. Walde, and T. Oberholzer, Curr.
Op. Coll. Int. Sci. {\bf 4}, 33 (1999).

\bibitem{clusters} P. Szab\'o, I. Scheuring, T. Cz\'ar\'an, and E. Szathm\'ary, Nature {\bf 420}, 340 (2002).

\bibitem{Damer2015} B. Damer and D. Deamer, Life {\bf 5}, 872 (2015).

\bibitem{Baaske2007} P. Baaske, F. M. Weinert, S. Duhr, K. H. Lemke,
M. J. Russell, and D. Braun, Proc. Natl. Acad.
Sci. U.S.A. {\bf 104}, 9346 (2007).

\bibitem{Matsumura2016} S. Matsumura, A. Kun, M. Ryckelynck, F. Coldren,
A. Szil\'agyi, F. Jossinet, C. Rick, P. Nghe,
E. Szathm\'ary, and A. D. Griffiths, Science {\bf 354},
1293 (2016).

\bibitem{SM} See Supplemental Material for the determination
of the parameter $\Lambda$ from the data of ref. [14], a
discussion of a stochastic version of the present
model and of other details concerning the phase
diagram.

\bibitem{Spiegelman1965} S. Spiegelman, I. Haruna, I. B. Holland, G. Beaudreau,
and D. Mills, Proc. Natl. Acad. Sci. USA {\bf 54}, 919
(1965).

\bibitem{Zadorin2017} A. S. Zadorin and Y. Rondelez, ArXiv e-prints
(2017), arXiv:1707.07461 [q-bio.PE].

\bibitem{Fisher1930} R. Fisher, The Genetical Theory of Natural Selection (Clarendon Press, Oxford, 1930).

\bibitem{Sneppen2014} K. Sneppen, Models of life (Cambridge University Press, 2014).
\end{thebibliography}
\end{document}